\documentclass[aps,twocolumn]{revtex4}
\usepackage{epsfig}
\usepackage{amsmath,amssymb}

\parskip=\medskipamount

\newcommand{\eq}[1]{(\ref{#1})}
\newcommand{\fig}[1]{Fig.\ref{#1}}

\newcommand{\be}{\begin{equation}}
\newcommand{\ee}{\end{equation}}

\newcommand{\barr}{\begin{array}}
\newcommand{\earr}{\end{array}}

\newcommand{\beqn}{\begin{eqnarray}}
\newcommand{\eeqn}{\end{eqnarray}}

\newcommand{\bs}{\begin{subequations}}
\newcommand{\es}{\end{subequations}}

\newcommand\disp{\displaystyle}

\newcommand{\la}{\left<}
\newcommand{\ra}{\right>}

\begin{document}

\title{Longest increasing subsequence as expectation of a simple nonlinear stochastic PDE with a
low noise intensity}

\author{E. Katzav$^1$, S. Nechaev$^2$\footnote{Also at: P.N. Lebedev
Physical Institute of the Russian Academy of Sciences, 119991, Moscow, Russia}, O. Vasilyev$^3$}
\affiliation{$^1$Laboratoire de Physique Statistique de l'Ecole Normale Sup\'{e}rieure, CNRS UMR
8550, 24 rue Lhomond, 75231 Paris Cedex 05, France \\
$^2$LPTMS, Universit\'e Paris Sud, 91405 Orsay Cedex, France \\
$^3$Department of Inhomogeneous Condensed Matter Theory,
Max-Planck-Institute f\"ur Metallforschung, Heisenbergstrasse 3,
D-70569 Stuttgart, Germany}

\date{\today}

\begin{abstract}
We report some new observation concerning the statistics of Longest Increasing Subsequences (LIS).
We show that the expectation of LIS, its variance, and apparently the full distribution function
appears in statistical analysis of some simple nonlinear stochastic partial differential equation
(SPDE) in the limit of very low noise intensity.
\end{abstract}

\maketitle

\section{Introduction}
\label{sect:1}

Last decade is marked by a breakthrough in the solution of a
classical probabilistic problem, known as "Ulam problem" discussed
in the mathematical literature over the years \cite{Ulam}. The
general setting of the Ulam problem is as follows. Take the unit
interval $[0,1]$ and pick up from it one after another $N$ random
numbers ($N\gg 1$) with uniform probability distribution. Having the
sequence of $N$ random numbers, one can extract from it the longest
increasing subsequence (LIS) of $k$ elements. The entries of this
subsequence are not obliged to be the nearest neighbors. There are
two basic questions of interest: i) what is the expectation $\la
k(N)\ra$ of LIS; and ii) what are the fluctuations of the mean
length of LIS. The first question has been answered in 1975 by
Vershik and Kerov \cite{VK} who have shown that $\la k(N)\ra =
2\sqrt{N}$ for $N\gg 1$. They derived this result by mapping the
expectation of the LIS to the expectation of the first row of
ensemble of Young tableaux over the Plancherel measure. If the
initial random sequence contains repeated numbers, then the first
line of the Young tableau corresponds to the longest {\it
nondecreasing} subsequence.

The progress in treatment of fluctuations of LIS was the subject of
recent investigations
\cite{tracy,deift,johannson,okunkov,aldous,spohn}. In these works an
exact asymptotic form of a properly normalized full limiting
distribution has been established (the so-called Tracy--Widom
distribution) of the first row of the Young tableau over the
Plancherel measure by mapping to the distribution of the largest
eigenvalues of some classes of random matrices \cite{tracy}. In
particular it has been shown that the properly normalized variance
of LIS has for $N\gg 1$ the asymptotic scaling form ${\rm Var}
[k(N)] \equiv \la k^2(N)\ra - \la k(N)\ra^2 = c N^{1/3}$, where $c =
\la [\chi - \la \chi \ra]^2 \ra = 0.8132...$ and $\chi$ has the
Tracy--Widom distribution for the Gaussian Unitary Ensemble.

We would like also to point out a deep connection between the
distributions of edge states of random matrices and of statistical
characteristics of some random growth models. In one of the most
stimulating work \cite{johannson}, it has been shown that a
(1+1)--dimensional model of directed polymers in random environment,
which is in the KPZ universality class, has the Tracy--Widom
distribution for the scaled height (energy). Around the same time
the authors of \cite{spohn} have found an exact mapping between a
specific polynuclear growth (PNG) model and the LIS problem. The
same Tracy--Widom distribution was reported in another class of
(1+1)--dimensional growth models called "oriented digital boiling"
model \cite{GTW}.

In our work we report some new observation concerning the statistics
of longest increasing subsequences. Namely, we show that the
expectation of LIS, its variance, and apparently the full
distribution function appears in a statistical analysis of some
properly scaled simple nonlinear stochastic partial differential
equation (SPDE) in a limit of very low noise intensity.

The paper is organized as follows. In Section \ref{sect:2} we
introduce all the necessary definitions. In particular, we describe
LIS as the hight profile of some uniform discrete growth problem
with nonlocal long--ranged interactions. In Section \ref{sect:3} we
consider the discrete short--ranged asymmetric uniform growth
process and derive the corresponding continuous space--time SPDE for
the height profile. In Section \ref{sect:4} we show that the
properly scaled limit of an infinitely small noise in the derived
SPDE adequately describes the growth with long--ranged interactions,
i.e. the statistics of LIS. The discussion and conclusions are
collected in Section \ref{sect:5}.

\section{LIS as a height profile in a uniform asymmetric non-local growth process}
\label{sect:2}

\begin{figure}[ht]
\epsfig{file=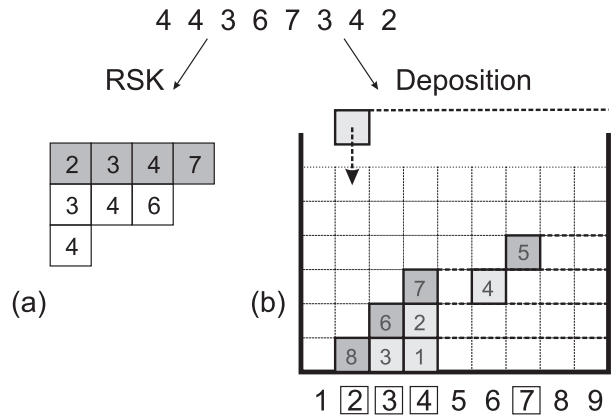,width=8cm} \caption{(a) Young tableau; (b) Asymmetric Ballistic
Deposition.}
\label{fig:1}
\end{figure}

The standard construction of a Young tableau for the set of real numbers is realized via the well
known Robinson--Schensted--Knuth algorithm (RSK) which can be found in many textbook on
representation theory, and for example, in \cite{rsk}. The RSK algorithm ensures that the first row
of the corresponding Young tableau would be the Longest Increasing Subsequence (LIS) for a given
set of numbers. To be specific, consider the example of $N=8$ numbers: $4,4,3,6,7,3,4,2$ taken at
random with uniform probability distribution from the support $\{1,...,9\}$. Following the RSK
algorithm of the Young tableau construction, we get the Young tableau shown in \fig{fig:1}a. Let us
describe now more geometrically obvious construction of LIS which refers to the discrete uniform
asymmetric ballistic deposition of elementary cells with long--ranged interactions. This
construction has appeared for the first time in \cite{maj_nech1}.

Consider the (1+1)D model of ballistic deposition in which the
columnar growth occurs sequentially on a linear substrate consisting
of $L$ columns with free boundary conditions. The time $t$ is
discrete and increases by $1$ with every deposition event. We start
with the flat initial condition, i.e., an empty substrate at $t=0$.
At any stage of the growth, a column (say the column $m$) is chosen
at random with probability $p=\frac{1}{L}$ and a "cell" is deposited
there which increases the height of this column by one unit: $h_m\to
h_m+1$. Once this "cell" is deposited, it screens all the sites at
the same level in all the columns to its right from future
deposition, i.e. the heights at all the columns to the right of the
column $m$ must be strictly greater than or equal to $h_m+1$ at all
subsequent times. Formally such a growth is implemented by the
following update rule. If the site $m$ is chosen at time $t$ for
deposition, then \be h_{m}(t+1)={\rm max}\{h_m(t), h_{m-1}(t),
\dots, h_{1}(t)\} + 1 \label{eq:1} \ee The model is anisotropic and
long--ranged and evidently even the average height profile $\la
h_m(t)\ra$ depends nontrivially on both the column number $k$ and
time $t$. Let us demonstrate now the a bijection between the longest
nondecreasing subsequence in the sequence of $N$ random numbers
uniformly taken from the support $\{1,2,3,...L\}$ and the height in
the uniformly growing heap with anisotropic infinite--ranged
interactions in a bounding box containing $L$ columns.

This bijection is defined by assigning the first line in the Young tableau to the "most top" (or
"visible" from the left--hand side) blocks---see the \fig{fig:1}b. For the configuration shown in
\fig{fig:1} we have the sequence of $N=8$ numbers: $4,4,3,6,7,3,4,2$ taken at random from the set
$\{1,...,9\}$ (there are $L=9$ columns in the box). The "visible" blocks define the longest
nondecreasing subsequence (LNS): $2,3,4,7$. To be exact, if there are few LNSs, our construction
extracts one of them -- exactly as in the RSK scheme.

As one can see in \fig{fig:1}, the first line in the Young tableau \eq{eq:1} exactly matches the
"most top" subsequence defined in our model and the distribution of the maximal height of a heap
coincides with the distribution of the longest nondecreasing subsequence.

\section{Short--ranged asymmetric ballistic growth and its continuous limit}
\label{sect:3}

\begin{figure}[ht]
\epsfig{file=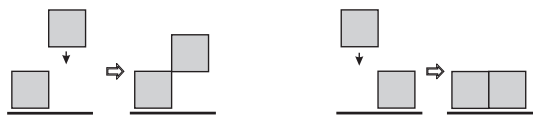, width=7cm} \caption{Updating rules for
the asymmetric NNN ballistic deposition defined in Eq.\eq{eq:2}.}
\label{fig:2}
\end{figure}

Consider a one--dimensional discrete model of ballistic deposition
with asymmetric (one--sided) next--nearest--neighboring (NNN)
interactions \bs \be h_j(t+1) = \max\{h_{j}(t), h_{j-1}(t)\} +
\eta_j(t) \label{eq:2}
\ee
where $h_j(t)$ is the height of a surface
at the lattice position $j$ and time $t$ (the time is discrete as
well) and $\eta_j(t)$ is a telegraph--like uncorrelated noise:
\be
\eta_j(t) = \left\{\begin{array}{ll} 1 & \mbox{with probability $p$} \medskip \\
0 & \mbox{with probability $1-p$}
\end{array}\right.
\label{eq:2b}
\ee
where
\be
\begin{cases} \overline{\eta_j(t)} = p \medskip  \\ \overline{\Big(\eta_j(t_1) -
\overline{\eta_j(t_1)} \Big)\,
\Big(\eta_m(t_2) - \overline{\eta_m(t_2)} \Big)} = p(1-p)\, \delta_{j,m}\, \delta_{t_1,t_2}
\end{cases}
\label{eq:3} \ee
\es
The bar denotes averaging over the distribution
\eq{eq:2b}. Equations \eq{eq:2}--\eq{eq:3} completely describe the
updating rules for the NNN discrete ballistic deposition with the
asymmetric interactions shown in \fig{fig:2}.

In order to derive a continuous model corresponding to \eq{eq:2}, We represent the $\max\{...\}$
operator in \eq{eq:2} at time $t$ by a sign--function ${\rm sign}[z]$:
\be
\barr{rl} \max\{h_{j-1},h_j\}= & \frac{1}{2} h_{j-1}\left(1+ \mbox{sign}\left[h_{j-1}-h_j\right]
\right) \medskip \\  + & \frac{1}{2}h_j\left(1+\mbox{sign}\left[h_j-h_{j-1} \right] \right) \earr
\label{eq:4}
\ee
where
$$
\mbox{sign}[z]=\begin{cases} +1 & \mbox{for $z>0$} \\ -1 & \mbox{for $z<0$} \end{cases}
$$
Substituting \eq{eq:4} into \eq{eq:2} we get after some simple algebra
\be
\barr{l} h_j(t+1)-h_j(t) = -\frac{1}{2}(h_j(t)-h_{j-1}(t)) \medskip \\ \hspace{0.5cm} +
\frac{1}{2}(h_j(t)-h_{j-1}(t)) \mbox{sign}[ h_j(t)-h_{j-1}(t)] + \eta_j(t) \earr
\label{eq:5}
\ee
In the last expression one can easily identify the finite--difference derivatives. Denoting the
spatial and temporal increments by $\Delta x$ and $\Delta t$ and taking into account that
$\mbox{sign}[az] = \mbox{sign}[z]$ (for $a>0$), we arrive at the following expression
\be
\barr{rl}
\partial_t h(x,t)= & -\frac{1}{2}\frac{\Delta x}{\Delta t} \partial_x h(x,t) \left(1-\mbox{sign}
\left[\partial_x h(x,t)\right] \right) \medskip \\ & + \frac{1}{\Delta t} \eta(x,t) \earr
\label{eq:6}
\ee
where $\partial_t h(x,t)$ and $\partial_x h(x,t)$ denote the partial derivatives of $h(x,t)$ with
respect to $t$ and $x$ correspondingly. Setting in \eq{eq:6} $\Delta t = \Delta x =1$ we arrive at
the nonlinear stochastic partial differential equation (SPDE) which is a continuous analog of the
discrete Asymmetric Ballistic Deposition described by \eq{eq:2}. The noise in \eq{eq:6} is still
defied by \eq{eq:2b}--\eq{eq:3}.

It is known \cite{katzav} that symmetric Ballistic Deposition model with NNN interactions in the
continuous limit has some relevance to KPZ equation \cite{kpz}. For better comparison of the KPZ
equation with our SPDE one, rewrite \eq{eq:6} as follows: \bs
\be
\partial_t h(x,t) = \frac{1}{2} \left|\partial_x h(x,t) \right| - \frac{1}{2} \partial_x h(x,t)
+ \eta(x,t)
\label{eq:7a}
\ee
or, equivalently,
\be  \partial_t h(x,t) = \left\{ \begin{array}{ll}
\disp \eta(x,t)-\partial_x h(x,t) & \mbox{if $\quad \disp \partial_x h(x,t)<0$}
\medskip \\ \disp \eta(x,t) & \mbox{if $\quad \disp \partial_x h(x,t)> 0$}
\end{array}
\right. \label{eq:7b} \ee \es with $\eta(x,t)$ given by
\eq{eq:2b}--\eq{eq:3}. One sees that the stochastic equation
\eq{eq:7a} does not contain a diffusion term and is very different
from the KPZ one. However, as will be shown below, the expectation
$\la h(x,t)\ra$ of Eq.\eq{eq:7a} converges for $t\gg 1$ and $x \gg
1$ to the expectation of LIS for a noise $\eta(x,t)$ with very small
intensity $p$ ($p< L^{-2}$). The same is valid for the variance,
$\mbox{Var}[h(x,t)]$, which demonstrates the KPZ scaling behavior.
All that allows us to conjecture that the full distribution of
$h(x,t)$ for small $p$ converges to a Tracy--Widom distribution of
LIS. Let us note that the structure of Eq.\eq{eq:7b} "ideologically"
resembles the structure of the one--dimensional Barenblatt model for
the field $u(x,t)$ with different diffusion constants for
$\partial_t u(x,t)>0$ and $\partial_t u(x,t)<0$ \cite{barenblatt}.

\section{SPDE in the infinitely rare noise regime and LIS}
\label{sect:4}

In this Section we show that the limit of a telegraph--like noise with small intensity in the
stochastic partial differential equation \eq{eq:7a} for the "height" function $h(x,t)$ has LIS
statistics. To do that, it is convenient to describe LIS in terms of (1+1)D Ballistic Deposition
(BD) set by Eq.\eq{eq:1}.

Let us begin with the limiting case of absence of any noise in
\eq{eq:7a}. The noiseless equation \eq{eq:7a} is invariant under the
scaling transformation $h(x,t)\to h(ax,at)$ for any $a$. Rewriting
\eq{eq:7a} in a finite--difference form on the lattice, we have: \be
\barr{rl} h(x,t+\Delta t) = h(x,t) & + \frac{1}{2}\Delta t
\left|\frac{h(x,t)- h(x-\Delta x,t)}{\Delta x}\right| \medskip \\ &
- \frac{1}{2}\Delta t  \left(\frac{h(x,t)-h(x-\Delta x,t)}{\Delta
x}\right) \earr \label{eq:8} \ee We can set $\Delta t=\Delta x=1$
and $x=j$ on the interval $0\le j \le L$ without any loss of
generality. The noiseless system defined by Eq.\eq{eq:8} has a
characteristic time scale $\tau \sim L$ of reaching the equilibrium.
Hence, after $t$ time steps, where $t>\tau$, the system described by
Eq.\eq{eq:8} reaches its stationary state generating a nondecreasing
staircase--like profile from any initial state. This can be seen
recursively: \be \left\{\barr{ll} h(j,t+1) = h(j-1,t) & \mbox{if
$h(j-1,t)>h(j,t)$}
\medskip \\ h(j,t+1) = h(j,t) & \mbox{if $h(j-1,t)<h(j,t)$} \earr \right.
\label{eq:9} \ee Since \eq{eq:9} is valid for any $0\le j \le L$,
after $t\sim L$ time steps, if in the initial state
$h(j=L,t=0)<h(j=1,t=0)$ then the heights at $x=L$ and $x=1$
equalize: $h(j=L,t)=h(j=1,t)$. This signals the appearance of
effective long--ranged spatial correlations in the system on
characteristic time scales of $\tau\sim L$.

To show the equivalence of SPDE and LIS in a noisy regime with small
intensity, it is convenient to introduce the enveloping functions
$r_m(t)$ and $r(x,t)$ for long--ranged (LR) BD and SPDE
correspondingly: \be \left\{\barr{rcll} r_{m}(t) & = &
\max\{h_{m}(t),\dots, h_{1}(t) \} & \mbox{LR BD (Eq.\eq{eq:1}})
\medskip \\ r(x,t) & = & \disp \max_{0 \le y \le x } \{h(y,t)\} & \mbox{SPDE}
\end{array} \right.
\label{eq:10}
\ee
If the intensity, $p$, of the noise $\eta(x,t)$ is small, after each noise event the function
$h(x,t)$ has time to reach its relaxed form $r(x,t)$ before the subsequent noise contribution is
added. The corresponding intensity, $p$, can be easily estimated. Namely, the mean time interval,
$\tau$, between positive noise events ($\eta(x,t)=1$) for a system of size $L$ and noise intensity
$p$ is $\tau \sim \frac{1}{L p}$. In more details this question is discussed in the Appendix. Since
the function $h(x,t)$ reaches the relaxed state $r(x,t)$ at $\tau>L$, we can estimate $p$ as
$p<L^{-2}$. Now Eqs.\eq{eq:1} and \eq{eq:7a} can be rewritten in the unified form:
\be
\left\{\barr{rcll}
h_{m}(t+1) & = & r_{m}(t)+1 & \mbox{long--ranged BD Eq.\eq{eq:1}} \medskip \\
h(x,t+\Delta t) & = & r(x,t) + \eta(x,t) & \mbox{SPDE with $p<L^{-2}$} \earr
\label{eq:11} \right.
\ee
In this form the equivalence between long--ranged BD and SPDE in a low--intensity noise regime is
clearly seen.

Thus, relying on the equivalence between statistics of LIS and low--intensity solutions of SPDE
\eq{eq:7a}, and knowing explicitly all the moments of the distribution of LIS (see, for example,
\cite{johannson}), we get the following asymptotic expressions for the expectation, $\la h(x,t)
\ra$, and for the variance, ${\rm Var}[h(x,t)] \equiv \la h(x,t)^2\ra - \la h(x,t)\ra^2$:
\beqn
\la h(x,t) \ra =  t p + 2 \sqrt{x t p} \label{eq:12a} \medskip \\
{\rm Var}[h(x,t)] = c (x t p)^{1/3} \label{eq:12b} \eeqn where
$h(x,t)$ is the asymptotic solution of the equation \eq{eq:7a} with
the noise distribution \eq{eq:3}. In \eq{eq:12b} $c = \la [\chi -
\la \chi \ra]^2 \ra = 0.8132...$ and $\chi$ has the Tracy--Widom
distribution for Gaussian Unitary Ensemble. Note that for
$p=\frac{1}{L}$ and $x=t=L$ ($L\gg 1$) we arrive in \eq{eq:12a} at
the Vershik--Kerov result for the expectation of LIS, $\la h(L) \ra
= 2 \sqrt{L}$. To be rigorous, this expression is beyond our
approximation since we are limited by the noise intensity $p_{\rm
cr} = \frac{1}{L^2}$. So, it is more convenient to work with the
general expressions \eq{eq:12a}--\eq{eq:12b}.

\begin{figure}[ht]
\epsfig{file=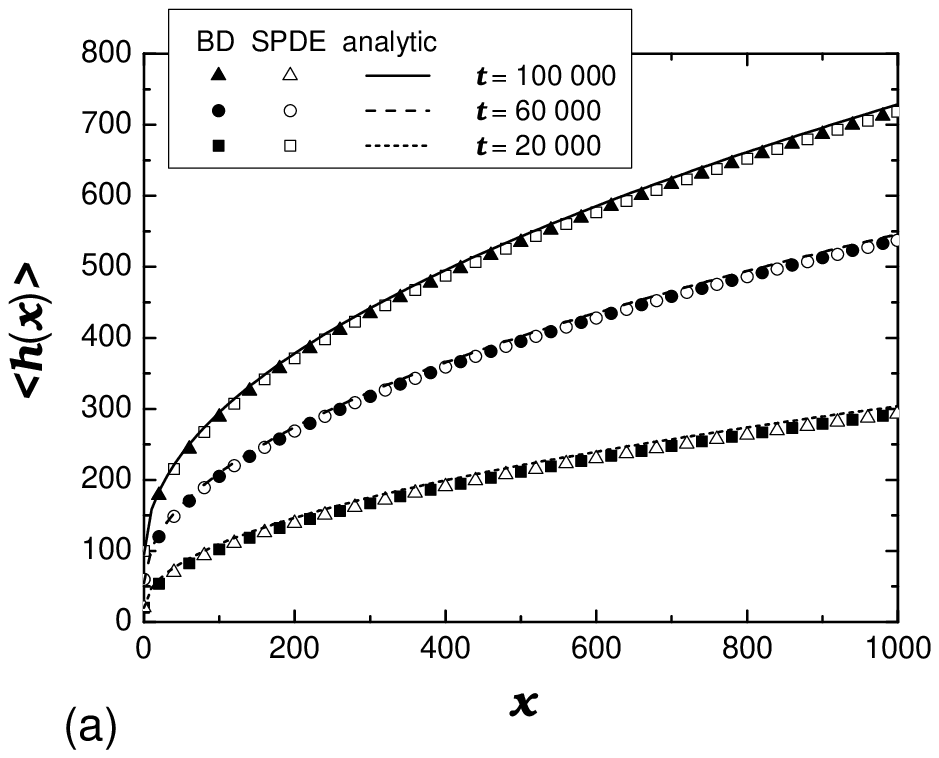,width=7cm} \\
\epsfig{file=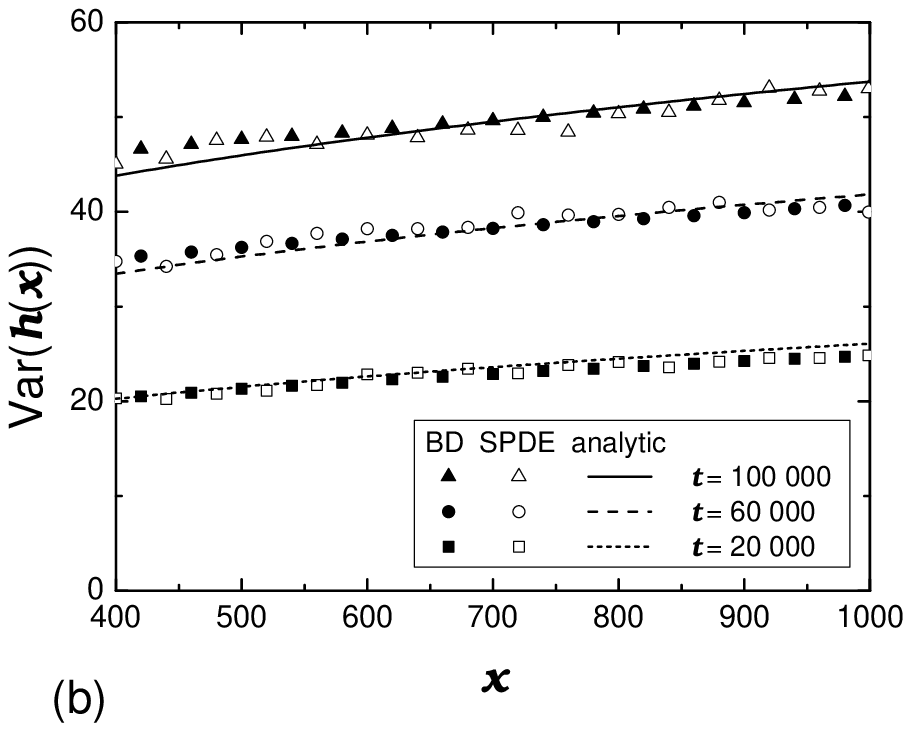,width=7cm} \caption{(a) The mean profile $\la h(x,t) \ra$ as a
function of $x$  for time points $t=2\times 10^4, 6\times 10^4, 10^5$ for BD and SPDE in comparison
with the analytical results \eq{eq:12a}; (b) The variance ${\rm Var}(h(x,t))$ as a function of $x$
at time points $t=2\times 10^4, 6\times 10^4, 10^5$ for BD and SPDE in comparison with the analytic
result \eq{eq:12b}. The length of the system here is $L=1000$.}
\label{fig:3}
\end{figure}

We confirm our analytic predictions \eq{eq:12a}--\eq{eq:12b} by numerical computations, simulating
directly the ballistic deposition process \eq{eq:1} and the stochastic partial differential
equation \eq{eq:7a} with noise intensity $p=\frac{0.1}{L}$ and initial condition $h(x,t=0)=0$. Note
that the intensity of the noise in our numerical simulations is $\frac{L}{10}$ times larger than
the upper estimate $p\sim \frac{1}{L^2}$, however the agreement with theoretical results
\eq{eq:12a}--\eq{eq:12b} is still very good. In particular, we plot in Fig.\ref{fig:3}a the average
profiles $\la h(x,t)\ra$ for long--ranged BD and SPDE as a function of $x$ ($0\le x \le L$) for $L
= 10^3$ at time points $t=2\times 10^4, 6\times 10^4, 10^5$. In the same figure we plot the
analytic expression \eq{eq:12a}. In Fig.\ref{fig:3}b we plot the variance ${\rm Var} (h(x,n)) =\la
h(x,t)^{2}\ra- \la h(x,t)\ra^{2}$ as a function of $x$ for the same time points $t$. It should be
noted that due to the numerical procedure of the definition of the height, the numerical
computation of the variance leads to the following interpolating expression for ${\rm
Var}[h(x,t)]=a(t) + c(x p t)^{1/3}$. However for $x\gg 1$ and $t={\rm const}$ we arrive
asymptotically at the expression \eq{eq:12b}. That is why we compare the expression \eq{eq:12b}
with the numerical data for large $x$ only where the term $a(t)$ is negligible. Moreover, the
agreement between long--ranged BD and SPDE is very good for the expectation and the variance ---
see \fig{fig:2}a,b.

\section{Conclusion}
\label{sect:5}

We have shown by the sequence of mappings that the expectation and
the variance of the random profile $h(x,t)$, described by the
stochastic nonlinear partial differential equation \eq{eq:7a} in the
limit of a noise with very low intensity, coincides with the
expectation \eq{eq:12a} and the variance \eq{eq:12b} of the longest
nondecreasing subsequence of the sequence of random integers. This
statement is also confirmed numerically. On the basis of the
obtained results we conjecture that not only the first moments, but
the full probability distribution function of the random variable
$h(x,t)$ coincides with the Tracy--Widom distribution appearing in
largest eigenvalue statistics of ensembles of random matrices.

Let us note that our analysis of Eq.\eq{eq:7a} is a bit indirect. It would be very desirable to get
expectation and variance \eq{eq:12a}--\eq{eq:12b} of the random variable $h(x,t)$ directly from the
solution of the Fokker--Planck equation which corresponds to the Langevin equation \eq{eq:7a}.
However on this way we have met some difficulties. Namely, rewrite Eq.\eq{eq:7a} as follows
\be
\partial_t h(x,t) = \frac{1}{2} \left|\partial_x h(x,t) \right| - \frac{1}{2} \partial_x h(x,t)
+ \xi(x,t) + p
\label{eq:fp0}
\ee
where $\xi(x,t) = \eta(x,t)-p$; $\overline{\xi(x,t)}=0$; $\overline{\xi(x,t) \xi(x',t')} = 2D
\delta_{x,x'} \delta_{t,t'}$ and $D=\frac{1}{2}p(1-p)$. Now we can write the formal expression for
the Fokker--Planck variational equation for the function $W(h,t)$ (see, for example, \cite{hh}):
\be
\barr{rl} \frac{\partial W(h,t)}{\partial t} = & - \int dx \frac{\delta}{\delta h}
\left\{\left(\frac{1}{2}\left|\partial_x h \right| - \frac{1}{2}\partial_x h + p\right)
W(h,t)\right\} \medskip \\ &  + D \int dx \frac{\delta^2}{\delta h^2} W(h,t) \earr
\label{eq:fp1}
\ee
which in turn can be rearranged in a form of two coupled equations:
\be \barr{ll}
\frac{\partial W(h,t)}{\partial t} & =  \int dx \left(-p\frac{\delta W(h,t)}{\delta h} + D
\frac{\delta^2 W(h,t)}{\delta h^2}\right) \medskip \\ & - \left\{\barr{ll} \int d x\, \partial_x
h\; \frac{\delta W(h,t)}{\delta h} & \mbox{if $\partial_x h<0$}
\medskip \\ 0  & \mbox{if $\partial_x h>0$} \earr \right. \earr
\label{eq:fp3}
\ee
where we have taken into account that $\frac{\delta}{\delta h} [\partial_x h(x,t)] = 0$.

\begin{figure}[ht]
\epsfig{file=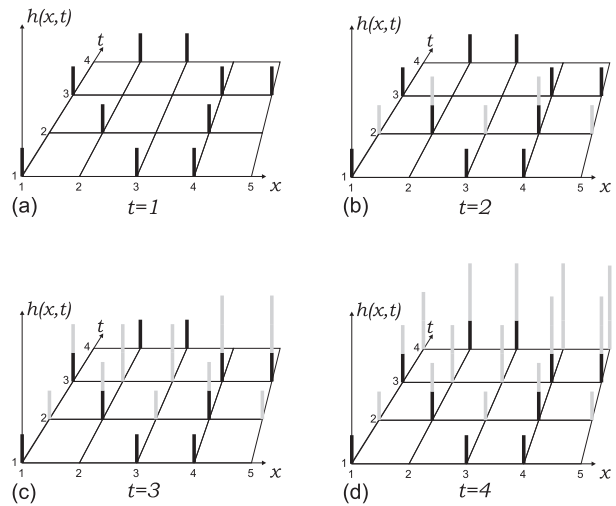,width=8cm} \caption{Visualization of the stochastic growth determined
by the recursion relation \eq{eq:13} for $x=1,...,5$ and for several time slices $t=1$ (a), $t=2$ (b),
$t=3$ (c), $t=4$ (d).}
\label{fig:4}
\end{figure}

The Fokker--Planck equation \eq{eq:fp3} corresponds to the stochastic process \eq{eq:7b} which can
be visualized alternatively as a (2+1)--dimensional correlated growth. Using the finite--difference
form \eq{eq:8} of \eq{eq:7b} on the lattice, where we set $\Delta x=\Delta t = 1$ and $x=j$, we
arrive at the following stochastic recursion relation
\be \barr{l} h(x,t+1) = h(x,t)+\eta(x,t) \medskip \\ \hspace{1cm} -\left\{ \barr{ll}h(x,t)-h(x-1,t)
& \mbox{if $h(x,t)<h(x-1,t)$} \medskip \\ 0 & \mbox{if $h(x,t)>h(x-1,t)$} \earr \right. \earr
\label{eq:13}
\ee
with the initial and boundary conditions $h(x,t=0)=h(x=0,t)=0$. Create now the initial
configuration distributing the random variable $\eta(x,t)$ in the plane $x\ge 1, t\ge 1$. Since
$\eta(x,t)$ takes the values 0 or 1, we can show $\eta=1$ by black unit segments as it is shown in
\fig{fig:4}. The initial configuration at $t=1$ is depicted in \fig{fig:4}a. Now applying the
equation \eq{eq:13} we can constrict recursively the configuration of the field $h(x,t)$ at
subsequent time moments. Few configurations of the field $h(x,t)$ for given initial distribution of
$\eta(x,t)$ is shown in \fig{fig:4}b-d for time moments $t=2,3,4$ and $x=1,...5$.

The (2+1)--dimensional stochastic growth shown in \fig{fig:4} bijectively corresponds to the
short--ranged ballistic growth considered in Section \ref{sect:3} and, hence, in the limit of small
intensity $p$ should correspond to the statistics of LIS. It would be very desirable to simulate
numerically directly the growth model shown in \fig{fig:4} to check the validity of the conjectured
bijection.

\acknowledgments

We are grateful to S. Majumdar for valuable discussions. The work is partially supported by the
grant ACI-NIM-2004-243 "Nouvelles Interfaces des Math\'ematiques" (France), and by EU PatForm
Marie Curie action (E.K.)

\begin{appendix}
\section{Equilibration time for SPDE}

In the limit of small $p$ ($p \ll 1$) the actual mean profile $\la h(x,t) \ra$  tends to the curve
$y(x,t) = t p+2 \sqrt{x t p }$. However for larger values of $p$ ($p \sim 1$) the resulting profile
does not have enough time to relax and hence is displaced below the curve $y(x,t)$. The question
which we address here concerns the estimation of the time interval, $\tau$, between the subsequent
deposition events sufficient to approach the stable solution $y(x,t)$.

We can estimate the equilibration time, $\tau$, from the following arguments. The slope of the mean
profile in the stationary state is
\be
\frac{\partial y(x,t)}{\partial x} =\sqrt{\frac{p t}{x}}
\ee
Hence, the average length of the horizonal "plateau" is about
\be
l(x) \simeq \frac{1}{\frac{\partial y(x,t)}{\partial x}}\simeq \sqrt{\frac{x}{pt}}
\label{eq:sat}
\ee
We define the equilibration time, $\tau$, as a time $t$ in \eq{eq:sat} during which the length
$l(x)$, of the plateau becomes that of the order of the system size, $L$. Substituting for $x$ in
\eq{eq:sat} the maximal value $x=L$, we arrive at the following estimate
$$
L \simeq  \sqrt{\frac{L}{p \tau}}
$$
what gives us
\be
\tau \simeq \frac{1}{L p}
\label{eq:tau}
\ee
Comparing the equilibration time \eq{eq:tau} with the characteristic time of the system relaxation,
$\tau \sim L$, we arrive at the estimate for the critical noise intensity, $p_{\rm cr}$,
$$
p_{\rm cr} \simeq \frac{1}{L^2}
$$
below which the stochastic partial differential equation \eq{eq:7a} has long--ranged behavior
typical for the LIS problem.

\end{appendix}


\begin{thebibliography}{99}

\bibitem{Ulam} S.M. Ulam, {\em Modern Mathematics for the Engineers}, ed. by
E.F. Beckenbach (McGraw-Hill: New York, 1961), p. 261

\bibitem{VK} A.M. Vershik and S.V. Kerov, Sov. Math. Dokl. {\bf 18}, 527
(1977)

\bibitem{tracy} C.A. Tracy and H. Widom, Comm. Math. Phys. {\bf 159}, 151 (1994);
{\bf 177}, 727 (1996); For a review see {\em Proceedings of the International Congress of
Mathematicians}, Beijing 2002, Vol. I, ed. Li Tatsien, Higher Education Press, Beijing 2002, p. 587

\bibitem{deift} J. Baik, P. Deift, and K. Johansson, J. Amer. Math. Soc. {\bf
12}, 1119 (1999)

\bibitem{johannson} K. Johansson, Comm. Math. Phys. {\bf 209}, 437 (2000)

\bibitem{okunkov} A. Okounkov, N. Reshetikhin, J. of Am. Math. Soc., {\bf 16}, 581 (2003)

\bibitem{aldous} For a review, see D. Aldous and P. Diaconis, Bull. Amer. Math.
Soc. {\bf 36}, 413 (1999)

\bibitem{spohn} M. Pr\"ahofer and H. Spohn, Phys. Rev. Lett. {\bf 84}, 4882
(2000); Physica A, {\bf 279}, 342 (2000)

\bibitem{GTW} J. Gravner, C.A. Tracy, and H. Widom, J. Stat. Phys. {\bf 102}, 1085 (2001)

\bibitem{rsk} C. Schensted, Canad. J. Math. {\bf 13} 179 (1961); D.E. Knuth,  Pacific J. Math. {\bf
34} 709 (1970)

\bibitem{maj_nech1} S.N.Majumdar, S.Nechaev, Phys.Rev.E, {\bf 69} 011103 (2004)

\bibitem{katzav} G. Costanza, Phys. Rev. E {\bf 55}, 6501 (1997); F.D.A. Aarao Reis, Phys. Rev.
E {\bf 63}, 056116 (2001); E. Katzav and M. Schwartz, Phys. Rev. E {\bf 70}, 061608 (2004)

\bibitem{kpz} M. Kardar, G. Parisi, and Y.-C. Zhang, Phys. Rev. Lett. {\bf 56}, 889 (1986)

\bibitem{barenblatt} G.I. Barenblatt, {\it Scaling phenomena in fluid mechanics}, (Cambridge Univ.
Press: Cambridge, 1994)

\bibitem{hh} T. Halpin-Healy, Y.-C. Zhang, Phys. Rep. {\bf 254}, 215 (1995)

\end{thebibliography}
\end{document}